# TRANSACTION-ORIENTED SIMULATION IN AD HOC GRIDS: DESIGN AND EXPERIENCE

Gerald Krafft and Vladimir Getov
Harrow School of Computer Science
University of Westminster
Watford Rd, Northwick Park, Harrow HA1 3TP, U.K.
E-mail: G.Krafft@gmx.net, V.S.Getov@wmin.ac.uk

**ABSTRACT**

In this paper we analyse the requirements of performing parallel transaction-oriented simulations within loosely coupled systems like ad hoc grids. We focus especially on the space-parallel approach to parallel simulation and on discrete event synchronisation algorithms that are suitable for transaction-oriented simulation and the target environment of ad hoc grids. To demonstrate our findings, a Java-based parallel simulator for the transaction-oriented language GPSS/H is implemented on the basis of the most promising shock-resistant Time Warp (SRTW) synchronisation algorithm and using the grid framework ProActive. The analysis of our parallel simulator, based on experiments using the Grid'5000 platform, shows that the SRTW algorithm can successfully reduce the number of rolled back transaction moves but it also reveals circumstances in which the SRTW algorithm can be outperformed by the normal Time Warp algorithm. Finally, possible improvements to the SRTW algorithm are proposed in order to avoid such problems.

## 1. INTRODUCTION

Transaction-oriented simulation is a special case of discrete event simulation that uses two types of objects, stationary objects called blocks and mobile objects called transactions that move through the model and change the state of the blocks. The movement of a transaction can be delayed or blocked by a stationary object but otherwise transactions always move at a certain simulation time meaning that the simulation time does not progress while a transaction is moved. Because such a movement of a transaction describes an action with a specific timestamp that changes the state of the model it is equivalent to an event in discrete event simulation. Inferred from this equivalency most aspects and findings for discrete event simulation can also be applied to transaction-oriented simulation and vice versa. Transaction-oriented simulation is best suited for the simulation of systems that are based on units of traffic competing for the use of specific resources which covers a wide area of applications. Typical examples include communication systems, transportation systems, manufacturing systems and general queuing systems. The best-known transaction-oriented simulation language is GPSS and its extended version GPSS/H (Schriber 1991).

Parallel and distributed computing offers a way to reduce the runtime of large and complex computer simulations as for instance required in engineering, military, biology and climate research. But the growing complexity of simulation models can reach the limits of today's high-performance parallel computer systems that in addition also induce a very significant cost factor.

Grid platforms can provide large-scale computing at lower costs by allowing several organisations to share their resources. But traditional grid infrastructures are relatively static environments that require a dedicated administrative authority and therefore are not well suited for transient short-term collaborations of small organisations with fewer resources. Ad hoc grids enable the design of such dynamic and transient resource-sharing infrastructures (Smith et al. 2004) that allow even small organisations or individual users to form grid environments on demand. They make grid computing and grid resources widely available to small organisations and mainstream users allowing them to perform resource demanding computing tasks like complex computer simulations.

There are several approaches to performing simulations distributed across a parallel computer system. The space-parallel approach (Fujimoto 1993) is one of these approaches that is robust, applicable to many different simulation types and can be used to speed up single simulation runs. It requires for the simulation model to be partitioned into relatively independent sub-systems that are assigned to logical processes (LPs). Each LP may then be processed on a different node. Synchronisation between these LPs is still required because the model sub-systems are usually not fully independent. A lot of past research has concentrated on different synchronisation algorithms for parallel simulation. Some of these are only suitable for certain types of parallel computers such as shared memory systems.

This work investigates the possibility of performing parallel transaction-oriented simulation in ad hoc grid environments with the main focus on the aspects of parallel simulation. Potential synchronisation algorithms and other simulation aspects are analysed in respect of their suitability for transaction-oriented simulation and ad hoc grids as the target environment and the chosen solutions are described and reasons for their choice given. To demonstrate the solutions a Java-based parallel simulator for the transaction-oriented language GPSS/H is implemented and evaluated using a set of simple example simulation models.

## 2. SYNCHRONISATION ALGORITHMS

### 2.1. Requirements

The choice of synchronisation algorithm can have a significant influence on how much of the parallelism that exists in a simulation model will be utilised by the parallel simulation system. An overview of the two main groups of synchronization algorithms described as conservative and optimistic algorithms can be found in (Das 1996). Conservative algorithms utilise the parallelism less well than optimistic algorithms because they require guarantees, which in most cases are derived from additional knowledge about the behaviour of the simulation model, like for instance the communication topology or lookahead attributes of the model. For this reason conservative algorithms are often used to simulate very specific systems where such knowledge is given or can easily be derived from the model. For general purpose simulation systems optimistic algorithms are better suited as they can utilise the parallelism within a model to a higher degree without requiring any guarantees or additional knowledge. The best-known optimistic algorithm is Time Warp (TW) (Jefferson 1985). But in many situations TW can show an over-optimistic behaviour leading to uncommitted simulation progress being undone as a result of rollbacks. Subsequent research has therefore focused on limiting the optimism in TW if required and in a self adapting way.

Another important aspect of choosing the right synchronisation algorithm is the relation between the performance properties of the expected parallel hardware architecture and the granularity of the parallel algorithm. In order for the parallel algorithm to perform well in general on the target hardware environment the granularity of the algorithm, i.e. the ratio between computation and communication has to fit the ratio of the computation performance and communication performance provided by the parallel hardware.

Considering the target environment of ad hoc grids and the goal of designing and implementing a general parallel simulation system based on the transaction-oriented simulation language GPSS/H we concluded that the best suitable synchronisation algorithm is an optimistic or hybrid algorithm that has a coarse grained granularity. The algorithm should require only little communication compared to the volume of computation it performs. At the same time the algorithm needs to be flexible enough to adapt to a changing environment, as this is the case in ad hoc grids. A further requirement is that the algorithm can be adapted to and is suitable for transaction-oriented simulation.

### 2.2. Algorithm Selection

A promising algorithm found for these requirements is the SRTW algorithm (Ferscha and Johnson 1999). This algorithm has some similarities with the elastic time algorithm (Srinivasan and Reynolds 1995) and also the adaptive memory management algorithm (Das and Fujimoto 1997) but at the same time is suitable for loosely coupled distributed systems like grids. Similar to the elastic time algorithm state vectors are used to describe the current states of all LPs plus a set of functions to determine the output vector but the SRTW algorithm does not require a global state. Instead each LP separately tries to optimise its parameters towards the best performance. Similar to the adaptive memory management algorithm the optimism is controlled indirectly be setting artificial memory limits but for the SRTW algorithm each LP will limit its own memory instead of using an overall memory limit for the whole simulator.

The SRTW algorithm is described by (Ferscha and Johnson 1999) as a fully distributed approach to controlling the optimism in TW that requires no additional communication between the LPs. It is based on the TW algorithm but extends each LP with a control component called logical process control component (LPCC) that constantly collects information about the current state of the LP using a set of sensors. These sets of sensor values are then translated into sets of indicator values representing state vectors for the LP. The LPCC will keep a history of such state vectors using a clustering technique so that it can search for past state vectors that are similar to the current state vector but provide a better performance indicator. An actuator value will be derived from the most similar of such state vectors that is subsequently used to control the optimism of the LP.

As the SRTW algorithm was designed for discrete event simulation its sensors and indicators had to be adapted to the equivalent values in transaction-oriented simulation.

## 3. SIMULATOR DESIGN ISSUES

### 3.1. End of Simulation

Another important aspect that had to be considered is the detection and correct handling of the simulation end. A transaction-oriented simulation is complete when the defined end state is reached, i.e. the termination counter reaches a value less or equal to zero. When using an optimistic synchronisation algorithm for the parallelisation of transaction-oriented simulation it is crucial to consider that optimistic algorithms will first execute all local events without guarantee that the causal order is correct. They will recover from wrong states by performing a rollback if it later turns out that the causal order was violated. Therefore, any local state reached by an optimistic LP has to be considered provisional until a global virtual time (GVT) message has been received that guarantees the state. In addition, it is most likely that at any point in real time each of the LPs has reached a different local simulation time so that after an end state has been reached by one of the LPs, which is guaranteed by a GVT, it is important to synchronise the states of all LPs. Thus, the combined end state from all model partitions is equivalent to the model end state that would have been reached in a sequential simulator.

The mechanism suggested here leads to a consistent and correct global end state of the simulation considering the problems mentioned above. For this mechanism the LP reaching a provisional end state is switched into the provisional end mode. In this mode the LP will stop to process any further transactions leaving the local model partition in the same state but it will still respond to and process control messages like GVT parameter requests and it will receive transactions from other LPs that might cause a rollback. The LP will stay in this provisional end mode until the end of the simulation is confirmed by a GVT or a received transaction causes a rollback with a potential re-execution that is not resulting in the same end state. While the LP is in the provisional end mode additional GVT parameters are passed on for every GVT calculation denoting the fact that a provisional end state has been reached and the simulation time and priority of the transaction that caused the provisional end. The GVT calculation process can then assess whether the earliest current provisional end state is guaranteed by the GVT. If this is the case then all other LPs are forced to synchronise to the correct end state by rolling back using the simulation time and priority of the transaction that caused the provisional end and the simulation is stopped.

## 3.2. Suitable Cancellation Technique

Transaction-oriented simulation has some specific properties compared to discrete event simulation. One of these properties is that transactions do not consume simulation time while they are moving from block to block. This has an influence on which of the synchronisation algorithms are suitable for transaction-oriented simulation but also on the cancellation techniques used. If a transaction moves from LP1 to LP2 then it will arrive at LP2 with the same simulation time that it had at LP1. A transaction moving from one LP to another is therefore equivalent to an event in discrete event simulation that when executed creates another event for the other LP with exactly the same timestamp. Because simulation models can contain loops, as it is for instance common for models of quality control systems where an item failing the quality control needs to loop back through the production process, this specific behaviour of transaction-oriented simulation can lead to endless rollback loops for certain cancellation techniques. Besides the original cancellation technique introduced by (Jefferson 1985) that is known as aggressive cancellation another cancellation technique called lazy cancellation was suggested by (Gafni 1985). Figure 1 compares the rollback behaviour of aggressive cancellation and lazy cancellation in respect of such a loop within the simulation model.

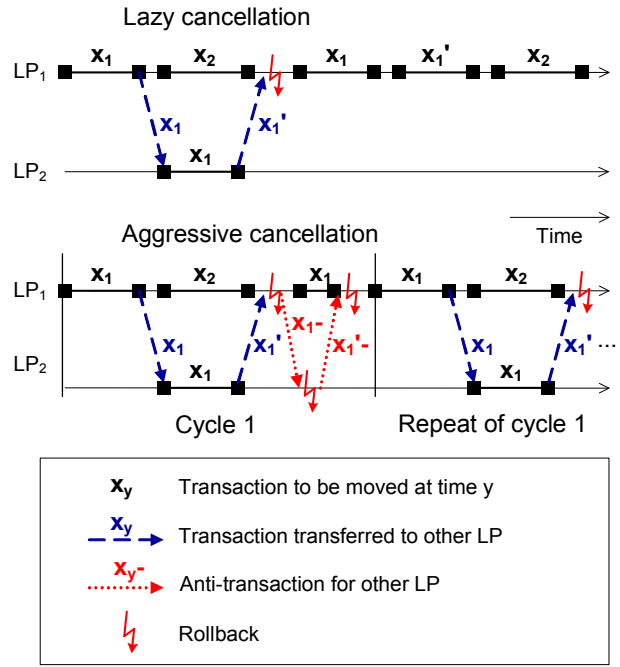

Figure 1: Cancellation in Transaction-Oriented Simulation

It shows the movement of transaction x1 from LP1 to LP2 but without a delay in simulation time the transaction is transferred back to LP1. As a result LP1 will be rolled back to the simulation time just before x1 was moved. At this point two copies of transaction x1 will exist in LP1. The first one is x1 itself which needs to be moved again and the second is x1' which is the copy that was send back from LP2. This is the point from where the execution differs between lazy cancellation and aggressive cancellation. In lazy cancellation x1 is processed again resulting in the same transfer to LP2. But because x1 was sent to LP1 already it will not be transferred again and no anti-transaction will be sent. From here LP1 just proceeds moving the transactions in its transaction chain according to their simulation time (transaction priorities are ignored for this example). Apposed to that in aggressive cancellation the rollback results in an anti-transaction being sent out for x1 immediately which causes a second rollback in LP2 and another anti-transaction for x1' being sent back to LP1. At the end both LPs will end up in the same state in which they were before x1 was processed by LP1. The same cycle of events would start again without any actual simulation progress.

In order to avoid the described endless rollback loops lazy cancellation needs to be used for parallel transaction-oriented simulation.

## 4. IMPLEMENTATION

The parallel transaction-oriented simulator was implemented using the Java-based grid environment ProActive (INRIA 2000) that is very well suited for ad hoc grids. The overall architecture of the parallel simulator follows the Master-Slave approach. Figure 2 shows the simplified architecture of the parallel simulator including its main components.

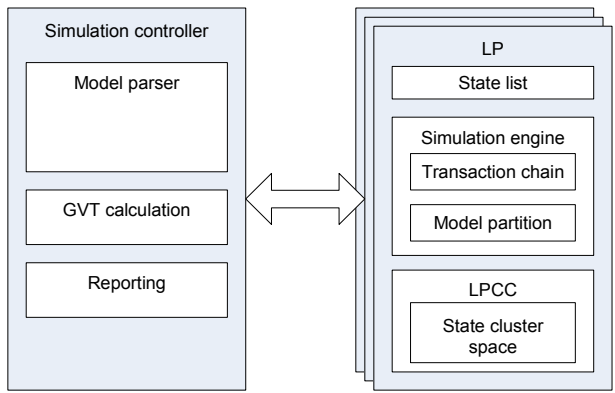

Figure 2: Architecture Overview

The main parts of the parallel simulator are the simulation controller and the LPs. The simulation controller steers the overall simulation. It is created when the user starts the simulation and will use the model parser component to read the simulation model file and parse it into an object structure representation of the model. After the model is parsed the simulation controller creates LP instances, one for each model partition. The simulation controller and the LPs are implemented as active objects of ProActive so that they can communicate with each other via remote method calls. Communication will take place between the simulation controller and the LPs but also between the LPs themselves, for instance, in order to exchange transactions. Note that the communication between the LPs is not illustrated in Figure 2. After the LPs have been created and initialised, they will receive the model partitions to simulate from the simulation controller and the simulation is started. The main component of each LP is the simulation engine, which contains the transaction chain and the model partition that is simulated. The simulation engine is the part that is performing the actual simulation. It is moving the transactions from block to block by executing the block functionality using the transactions. The implementation of the LPs follows the optimistic TW algorithm. It uses state checkpointing, i.e. each LP contains a state list that stores historic simulation states allowing them to perform rollbacks if required. Lazy cancellation is implemented in order to propagate the cancellation of transactions already sent to other LPs. There are other lists within the simulation engine that are not shown in Figure 2, for instance the list of transactions received and the list of transactions sent. The LPs are extended further to operate according to the SRTW algorithm as described by (Ferscha and Johnson 1999). This is achieved by adding an LPCC and specific sensors into each LP. The sensor values are periodically read by the LPCC and converted into indicators. The most important indicators for the SRTW algorithm are the number of events committed per second and the average number of events in use. The first one describes the simulation progress and the second is artificially limited by the LPCC in order to reduce the optimism if required. For transaction-oriented simulation the equivalent indicators are the number of committed transaction moves per second and the average number of uncommitted transaction moves consisting of transaction moves that have been performed but not yet committed and transaction moves that are scheduled to be performed. The LPCC stores each indicator set in a cluster space and then uses it to find a similar past indicator set that promises better performance and to derive a new actuator value from the indicator set found. An option to disable the LPCC allows the simulator to either operate in SRTW mode or in TW mode.

The simulation controller will perform GVT calculations necessary to establish the overall progress of the simulation and to allow LPs to reclaim memory through fossil collection. GVT calculation will also be used to confirm a provisional simulation end state that might be reached by one of the LPs.

When the end of the simulation is reached then the simulation controller will ensure that the partial models in all LPs are set to the correct and consistent end state and it will collect and assemble information from all LPs to output the post simulation report.

## 5. SIMULATION RESULTS

The performance of a distributed simulation depends on several factors such as the simulation model and how it is partitioned, the hardware performance of the nodes processing the LPs plus any additional loads on these nodes and the performance of the communication channels between the LPs. It also depends on how efficient the synchronization algorithm can utilise the parallelism within the model. This includes whether an optimistic simulation leads to many and possibly cascaded rollbacks because of over-optimistic processing. Over-optimistic processing can itself be a result of the simulation model or different and changing processing speeds of the LPs.

The evaluation of the simulator therefore concentrates on whether the SRTW algorithm can limit the optimism compared to TW if required and on other effects it has on simulation performance. The two example simulation models used were deliberately kept very simple in order to evaluate specific aspects of the algorithms. Each example simulation model was run once in SRTW mode and once in TW mode. The validation runs were performed on the Grid'5000 platform with each LP and the simulation controller using a separate node of the Azur cluster at the Sophia Antipolis site. This cluster consists of IBM eServer 325 machines with two 2.0GHz AMD Opteron 246 CPUs per node. Communication between the nodes took place through a gigabit Ethernet link.

### 5.1. Reduction of Rolled Back Transaction Moves

The simulation model used for the first evaluation is shown in Figure 3. It contains two partitions each simulated by a separate LP. Both partitions have a GENERATE block and a TERMINATE block but in addition partition 1 also contains a TRANSFER block that with a very small probability of 0.001 sends some of its transactions to partition 2. The whole model is constructed so that partition 2 is ahead of partition 1 regards simulation time, achieved through the different configuration of the GENERATE blocks, and that occasionally partition 2 receives a transaction from partition 1. Because partition 2 is ahead in simulation time, this will lead to rollbacks in partition 2. The simulation stops after 120000 transactions have been terminated in the second partition. This model attempts to emulate the common scenario where a distributed simulation uses nodes with different performance parameters or partitions that create different loads so that during the simulation the LPs drift apart and some of them are further ahead in simulation time than others leading to rollbacks and re-execution.

```
PARTITION Partition1,120000
GENERATE 1,0
TRANSFER 0.001,Label1
TERMINATE 0
PARTITION Partition2,120000
GENERATE 3,0,5000
Label1 TERMINATE 1
```

Figure 3: Simulation Model 1

The model was simulated once in SRTW mode and once in TW mode by enabling or disabling the LPCCs within all LPs of the simulator. We found that in SRTW mode the LPCC of LP2 successfully reduces the number of rolled back transaction moves compared to the TW mode by limiting the number of uncommitted transaction moves using the actuator. Figure 4 shows the actuator values set by the LPCC during the simulation. For some of the processing intervals no actuator value was set because the LPCC could not find a past state vector with a better performance indicator.

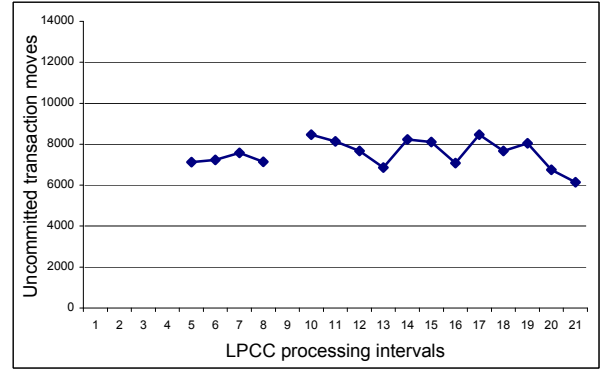

Figure 4: LP2 Actuator Value Graph

The actual reduction in the number of rolled back transaction moves within LP2 over the whole simulation can be seen in Table 1. It shows that the simulation run using SRTW required 74232 less rolled back transaction moves, which is around 19% less compared to the simulation run using TW. As a result the total number of transaction moves performed by the simulation was also reduced leading to a better average simulation performance of 5502 time unites per second in SRTW mode compared to 4637 time unites per second in TW mode.

Table 1: LP2 Processing Statistics

| LP statistic item | SRTW | TW |
| --- | --- | --- |
| Total transaction moves rolled back | 308790 | 383022 |
| Total simulated transaction moves | 428790 | 503022 |

### 5.2. TW Outperforming SRTW

During the testing of the parallel simulator we found that in some cases the TW algorithm can outperform the SRTW algorithm. This second evaluation demonstrates this in an example. The simulation model is very similar to the one used in the first evaluation. It contains two partitions with the first partition transferring some of its transactions to the second partition but this time the GENERATE blocks are configured so that the first partition is ahead in simulation time compared to the second. The simulation is finished after 5000 transactions have been terminated in one of the partitions. The complete simulation model can be seen in Figure 5.

```
PARTITION Partition1,5000
GENERATE 1,0,4000
TRANSFER 0.3,Label1
TERMINATE 1
PARTITION Partition2,5000
GENERATE 1,0
Label1 TERMINATE 1
```

Figure 5: Simulation Model 2

As a result of the changed GENERATE block configuration and the first partition being ahead of the second partition in simulation time, all transactions received by partition 2 from partition 1 are in the future for partition 2 and no rollbacks will be caused. But it will lead to a steady increase of the number of outstanding transactions within partition 2 pushing up the indicator for the number of uncommitted transaction moves during the simulation.

The first simulation run was performed with the LPCC enabled, i.e. in SRTW mode. The significant effect of the simulation run is that the LPCC in LP2 starts setting actuator values in order to steer the local simulation processing towards a past state that promises better performance but because the number of uncommitted transaction moves within the second partition increases as a result of the transactions received from partition 1 the actuator values set by the LPCC tend to be lower than the current number of uncommitted transaction moves resulting in the actuator value being exceeded and the LP being switched into cancelback mode. The cancelback mode forces the LP2 to temporarily stop processing transactions and to cancel back some of the transactions received from LP1 leading to the overall simulation progress being slowed down. Figure 6 shows the actuator values applied by the LPCC in LP2. For most of the simulation the actuator was set to a value of around 2000 uncommitted transaction moves. All intervals that had an actuator value set led to the LPCC switching the LP2 into cancel back mode resulting in a significantly reduced rate of committed transaction moves in LP2 as shown in Figure 7.

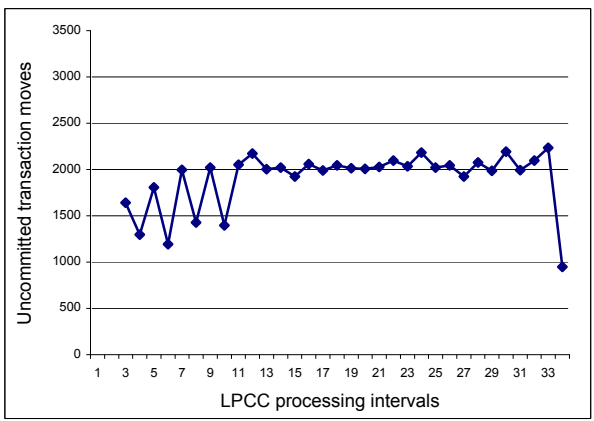

Figure 6: LP2 Actuator Value Graph

The second simulation run of this model was performed with the LPCC disabled, i.e. in TW mode. There were no rollbacks during the simulation and none of the LPs were artificially slowed down leading to an optimum average simulation performance for the model and setup of 165.7 time units per second. In SRTW mode the average simulation performance for the same model was dramatically reduced to 27.7 time units per second.

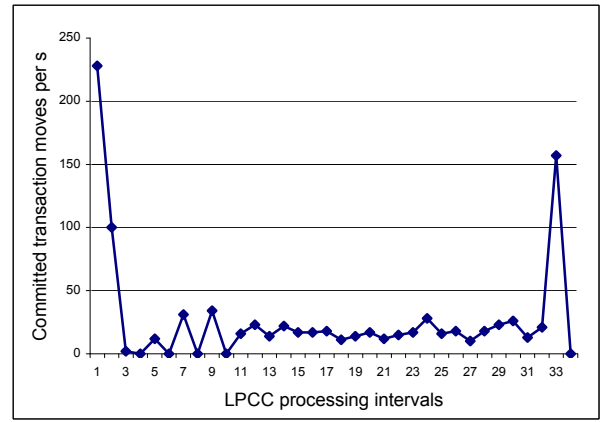

Figure 7: LP2 Simulation Progress Rate

## 6. DISCUSSION AND FUTURE WORK

The evaluation using the first example simulation model demonstrated that the SRTW algorithm can successfully reduce the amount of rolled back simulation work compared to TW. By limiting the optimism the SRTW algorithm reduced the number of rolled back transaction moves around 19%.

The second example simulation model revealed a problem of the SRTW algorithm. The TW algorithm provides ideal performance for this model because no rollbacks were caused and therefore none of the simulation work needed to be undone. Any superior synchronization algorithm would be expected to perform this simulation model with similar performance like TW. However, the SRTW algorithm performs significantly worse than the TW algorithm because it attempts to limit optimism in an LP that is already behind in respect of simulation time and that is not causing any rollbacks leading to a significant slowdown of the overall simulation progress. The SRTW algorithm fails in this scenario because the adaptive optimization within the LPs is purely based on local information that is not always sufficient for the algorithm to make the correct decision. Making the LPs aware of their position within the global progress window (GPW) as suggested by (Tay et al. 1997) could be a way of avoiding such problems. The SRTW algorithm already requires GVT calculations in order to establish the values of its performance indicators and this GVT calculation could easily be extended to calculate the global furthest time (GFT) giving all LPs an additional indicator of their position within the GPW.

In future work the parallel simulator could be extended to use a changed SRTW algorithm that is aware of the GPW and perhaps also implement the adaptive throttle scheme as described by (Tay et al. 1997) so that further comparisons of these algorithms are possible. Future studies could also investigate how the SRTW algorithm or an improved version of it reacts to communication delays that are likely to occur in Internet based grids.

# 7. CONCLUSION

We briefly discussed the requirements for a synchronisation algorithm suitable for ad hoc grid environments as well as transaction-oriented simulation. Further requirements for parallel transaction-oriented simulation were analysed and possible solutions suggested. The SRTW algorithm was chosen as a promising algorithm that fulfils the requirements. The algorithm was adapted to transaction-oriented simulation and a parallel simulator was implemented using the grid environment ProActive. Our parallel simulator can operate in SRTW mode as well as in TW mode allowing a comparison of the two algorithms for different transaction-oriented simulation models.

The evaluation of the parallel simulator showed that the SRTW algorithm can successfully reduce the number of rolled back transaction moves, which for simulations with many rollbacks will lead to a better simulation performance. But it also revealed a weakness of the SRTW algorithm. Because LPs try to optimise their properties based only on local information it is possible for the SRTW algorithm to perform significantly worse than the TW algorithm. Future work on this simulator could improve the SRTW algorithm by making the LPs aware of their position within the global progress of the simulation.

## ACKNOWLEDGMENT

This research work was funded partially by the European Commission under the research and development project GridCOMP (Contract IST-2006-034442). Experiments presented in this paper were carried out using the Grid'5000 experimental testbed, an initiative from the French Ministry of Research through the ACI GRID incentive action, INRIA, CNRS, RENATER and other contributing partners (https://www.grid5000.fr).

## AUTHOR BIOGRAPHIES

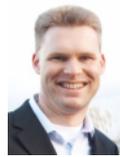

**GERALD KRAFFT** was born in Schwedt/Oder, Germany and first went to the University of Wismar, where he studied Computer Science and obtained his degree in 1998. He subsequently relocated to London, U.K. where he works as a senior software developer and joined the University of Westminster for a postgraduate degree in Advanced Computer Science, which he completed with distinction in 2007. He has a particular interest in parallel and distributed systems and computer simulation. His Web-page is http://perun.hscs.wmin.ac.uk/~gerald/.

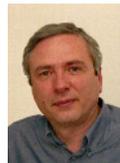

**VLADIMIR GETOV** leads the Distributed and Intelligent Systems Group at the University of Westminster in London. His current research interests are focussed on component-oriented design of Grid platforms and applications, autonomous distributed systems, parallel architectures and performance, mixed-language high-performance programming environments with Java, and hybrid programming models and paradigms. Professor Getov has over 100 publications including edited volumes, articles or chapters in books, journal and conference papers, technical reports, as well as invited and tutorial lectures, seminars, design prototypes, etc. His Web-page is http://perun.hscs.wmin.ac.uk/~vsg/.